\documentstyle[12pt,epsf]{article}
\newcommand{\dis}{\displaystyle}

\newcommand{\Dleftad}{{\overleftarrow D}_\alpha}

\newcommand{\DDmu}{\bigtriangledown_\mu}
\newcommand{\DUmu}{\bigtriangledown^\mu}

\begin{document}

\thispagestyle{empty}
\begin{flushright}
CERN-TH/98-209\\
LNF-98/021(P)\\
June 1998

\end{flushright}
\vspace*{1.5cm}
\centerline{\Large\bf The CP Conserving Contribution to 
                      $K_L\to\pi^0\nu\bar\nu$}
\vspace*{0.3cm}
\centerline{\Large\bf in the Standard Model}
\vspace*{2cm}
\centerline{{\sc Gerhard Buchalla${}^a$} and {\sc Gino Isidori${}^b$}}
\bigskip
\centerline{\sl ${}^a$Theory Division, CERN, CH-1211 Geneva 23,
                Switzerland}
\centerline{\sl ${}^b$INFN, Laboratori Nazionali di Frascati, 
                I-00044 Frascati, Italy}

\vspace*{1.5cm}
\centerline{\bf Abstract}
\vspace*{0.3cm}
\noindent 
The rare decay $K_L\to\pi^0\nu\bar\nu$ is dominated by direct 
CP violation and can be computed with extraordinarily high precision.
In principle also a CP conserving contribution to this process can
arise within the Standard Model. We clarify the structure of the
CP conserving mechanism, analyzing both its short-distance and
long-distance components. It is pointed out that the calculation of
the CP conserving amplitude, although sensitive in part to
nonperturbative physics, is rather well under control. The resulting
CP conserving contribution to the rate for $K_L\to\pi^0\nu\bar\nu$
turns out to be very strongly suppressed due to several factors,
which we discuss in detail.
\vfill

\newpage
\pagenumbering{arabic}

\section{Introduction}

The rare decay mode $K_L\to\pi^0\nu\bar\nu$ provides one of the
most promising opportunities in flavor physics. It proceeds through
a loop-induced flavor-changing neutral current (FCNC) transition
and therefore probes Standard Model (SM) dynamics at the quantum level.
In particular, $K_L\to\pi^0\nu\bar\nu$ offers a unique possibility to
test the mechanism of CP violation \cite{LIT}. The decisive virtue of
$K_L\to\pi^0\nu\bar\nu$ is the exceptional degree to which the
theoretical analysis of this decay is under control, with theoretical
uncertainties at the level of a few percent at most. The basic reason
for this favorable situation is the absence of contributions from
virtual photons, resulting in a power-like ($\sim m^2_i/M^2_W$, 
$i=u, c, t$) GIM cancellation pattern of the FCNC amplitude.
Reliably calculable contributions from high-mass intermediate states
($m_t\gg m_c\gg\Lambda_{QCD}$) are therefore systematically enhanced
over long-distance effects. This short-distance dominance is further 
reinforced in $K_L\to\pi^0\nu\bar\nu$, compared to the related 
mode $K^+\to\pi^+\nu\bar\nu$, by the large CP violating
phase associated with the top loops.

Although the detection of $K_L\to\pi^0\nu\bar\nu$ is very
challenging, due to a very small  branching fraction 
($\sim 3\cdot 10^{-11}$ within the SM) and a difficult
signature, considerable interest exists around the world in studying
this decay experimentally and important steps toward this goal have
already been undertaken. An experiment with the sensitivity to measure
$B(K_L\to\pi^0\nu\bar\nu)$ at the SM level has been proposed at
Brookhaven (BNL-E926) \cite{E926}. The KAMI collaboration at Fermilab
has published an Expression of Interest for such a measurement in the
Main Injector era \cite{KAMI} and plans to search for this decay 
with similar sensitivities also exist at KEK in Japan \cite{KEK}.
Finally, the potential of
KLOE at DA$\Phi$NE (the Frascati $\Phi$-Factory)
to search for $K_L\to\pi^0\nu\bar\nu$, with a smaller but still
interesting sensitivity and on a short time scale, has been
recently emphasized in \cite{BCI}. 

Let us briefly summarize the status of the theory of 
$K_L\to\pi^0\nu\bar\nu$. The relevant low-energy effective Hamiltonian
that describes the short-distance FCNC interaction inducing
$K_L\to\pi^0\nu\bar\nu$ can be written as
\begin{equation}\label{heff}
{\cal H}_{eff}=\frac{G_F}{\sqrt{2}}
\frac{\alpha}{2\pi\sin^2\Theta_W}\lambda_t X_0(x_t)\
(\bar sd)_{V-A}(\bar\nu_l\nu_l)_{V-A}+ {\rm h.c.}~,
\end{equation}
where $\lambda_i=V^*_{is}V_{id}$, $x_i=m^2_i/M^2_W$ and
\begin{equation}\label{x0x}
X_0(x)=\frac{x}{8}\left[\frac{x+2}{x-1}+\frac{3x-6}{(x-1)^2}
\ln x\right]~.
\end{equation}
The charm quark contribution, sizable in the charged mode
$K^+\to\pi^+\nu\bar\nu$, is completely negligible for 
$K_L\to\pi^0\nu\bar\nu$ and has been omitted from (\ref{heff}).
The Hamiltonian (\ref{heff}), with the one-loop function $X_0(x)$
first calculated in \cite{IL}, provides a good starting point for
the calculation of $K_L\to\pi^0\nu\bar\nu$ in the Standard Model.
Several important refinements have subsequently been added to the
theoretical analysis of this decay. The dominant uncertainty of the
lowest order prediction can be eliminated by including NLO QCD effects
\cite{BB12}. The hadronic matrix elements
$\langle\pi|(\bar sd)_V|K\rangle$ are known from the leading
semileptonic decay $K^+\to\pi^0e^+\nu$ using isospin symmetry.
Corrections due to small isospin breaking effects
have been computed in \cite{MP}. Finally, the impact of higher order
electroweak effects ($\sim G^2_F m^4_t$ in the amplitude) has been
studied in \cite{BB7}.

Overall the theoretical uncertainties in the $K_L\to\pi^0\nu\bar\nu$
branching fraction are thus under control to better than $\pm 3\%$,
assuming that potential long-distance effects can be neglected.
Such effects, which are not included in the description provided
by (\ref{heff}), have been estimated in \cite{RS} and were indeed
found to be safely negligible.

The dominant short-distance mechanism for $K_L\to\pi^0\nu\bar\nu$
based on (\ref{heff}) violates CP symmetry as a consequence of the
CP transformation properties of $K_L\approx K^0_{CP-odd}$, $\pi^0$
and the hadronic $(V-A)$ transition current
$\lambda_t (\bar sd)_{V-A}+\lambda^*_t (\bar ds)_{V-A}$. These imply
(in standard CKM phase conventions)
\begin{equation}\label{akl}
A(K_L\to\pi^0\nu\bar\nu)\sim {\rm Im}\lambda_t
 \langle\pi^0|(\bar sd)_{V-A}|K^0\rangle\ ,
\end{equation}
which would be zero in the limit of CP conservation. By contrast,
the long-distance effects studied in \cite{RS} survive in this limit. 
In this respect CP violation in $K_L\to\pi^0\nu\bar\nu$ differs from
the case of $K_L\to\pi\pi$, where the transition itself is forbidden by
CP invariance.

The purpose of this Letter is to present a systematic discussion
of the CP conserving contribution to $K_L\to\pi^0\nu\bar\nu$ in the
Standard Model. This question is of interest not only for estimating
theoretical uncertainties from long-distance dynamics, but also as a
matter of principle, in view of the role of $K_L\to\pi^0\nu\bar\nu$
as a CP violation ``standard''. In some  New Physics scenarios the 
CP conserving contributions to $K_L\to\pi^0\nu\bar\nu$ can be important
\cite{GN,KMT}. Thus also from this perspective it is interesting to
quantify the size of the CP conserving effect that, in principle,
exists in the Standard Model itself.

The present analysis confirms the estimate of \cite{RS} that the CP
conserving contribution to $K_L\to\pi^0\nu\bar\nu$ in the Standard
Model is very small. We differ, however, from \cite{RS} in our general
approach and include in our discussion in particular the short-distance
contribution to the CP conserving amplitude that has not been
considered before.

In Sections 2 and 3 we analyze, respectively, the short-distance 
and the long-distance mechanism of the CP conserving 
$K_L\to\pi^0\nu\bar\nu$ amplitude. We conclude in Section 4.

\section{The short-distance part of the CP conserving amplitude}

In the limit of exact CP symmetry, the leading term in the operator product
expansion (OPE) for $s\to d\nu\bar\nu$ transitions (\ref{heff})
gives a vanishing contribution to the $K_L\to\pi^0\nu\bar\nu$ amplitude.
More explicitly, in this limit the matrix element of the hadronic
$\Delta S=1$ transition current is given by 
\begin{equation}\label{psdk}
\langle\pi^0(p)|(\bar sd)_{V-A}+(\bar ds)_{V-A}|K_L(k)\rangle~,
\end{equation}
where the CKM parameters, chosen to be real, have been factored out. 
Using the CP transformation properties
($\tilde k_\mu\equiv k^\mu$)
\begin{equation}\label{cpp}
CP|K_L(k)\rangle=-|K_L(\tilde k)\rangle~, \qquad
CP|\pi^0(p)\rangle=-|\pi^0(\tilde p)\rangle~,
\end{equation}
\begin{equation}
CP(\bar sd)^\mu_{V-A}(CP)^{-1}=-(\bar ds)_{V-A,\mu}~,
\end{equation}
the matrix element (\ref{psdk}) is seen to be zero. 

\begin{figure}[t]
    \begin{center}
       \setlength{\unitlength}{1truecm}
       \begin{picture}(6.0,3.5)
       \epsfxsize 6.0 true cm
       \epsfysize 3.5 true cm
       \epsffile{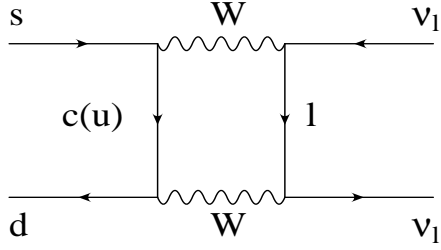}
       \end{picture} 
    \end{center}
    \caption{Box diagram which generates the leading 
short-distance CP conserving contribution to $K_L\to\pi^0\nu\bar\nu$.}
    \protect\label{figwbox}
\end{figure}

A nonvanishing
CP conserving contribution, though forbidden by (\ref{heff}), can however
arise at higher orders in the OPE. The leading effect of this type
comes from the $W$-box diagram with intermediate charm (up) quarks
depicted in Fig.~\ref{figwbox}. Matching this amplitude onto an effective
Hamiltonian leads to the following, CP conserving interaction term
of dimension eight
\begin{equation}\label{hcpc}
{\cal H}_{CPC}=-\frac{G_F}{\sqrt{2}}\frac{\alpha}{2\pi\sin^2\Theta_W}
\lambda_c\, \ln\frac{m_c}{\mu}\, \frac{1}{M^2_W}T_{\alpha\mu}
\bar\nu({\overleftarrow{\partial^\alpha}} -\partial^\alpha)\gamma^\mu
(1-\gamma_5)\nu~,
\end{equation}
\begin{equation}\label{tam}
T_{\alpha\mu}=\bar s{\overleftarrow D}_\alpha\gamma_\mu(1-\gamma_5)d-
\bar d\gamma_\mu(1-\gamma_5)D_\alpha s~.
\end{equation}
In this case we have $CP\, (T_{\alpha\mu})\, (CP)^{-1}=+T^{\alpha\mu}$,
using the same CP conventions as above, and
$\langle\pi^0|T_{\alpha\mu}|K_L\rangle$
is in general non-zero. Note that the relative minus sign in (\ref{tam})
results from the neutrino current in (\ref{hcpc}) being
antisymmetric ($\sim(q_1-q_2)^\alpha$) in the neutrino and
antineutrino momenta ($q_1$ and $q_2$ respectively), and the
hermiticity of ${\cal H}_{CPC}$.

In order to obtain the charm contribution to ${\cal H}_{CPC}$,
shown in (\ref{hcpc}), we have expanded the diagram of Fig.~\ref{figwbox}
in powers of external momenta divided by the charm quark mass $m_c$
(after contracting the $W$-boson lines).
Only the leading term, of second order in momenta
and independent of $m_c$ up to a logarithm, has been retained.
A similar expression would
hold for the up quark contribution if also $m_u\gg\Lambda_{QCD}$. Then
a finite coefficient would result as a consequence of the GIM
cancellation and the logarithms would combine to $\ln(m_c/m_u)$. In
reality $m_u$ is small and the up quark contribution has to be treated
nonperturbatively. We will come back to this point below. In the meantime
we have included in (\ref{hcpc}) only the charm part and kept the explicit
dependence on a renormalization scale $\mu$, which is to be canceled
by the up quark sector. For an order-of-magnitude estimate 
$\mu\;\raisebox{-.4ex}{\rlap{$\sim$}} \raisebox{.4ex}{$>$}\;\Lambda_{QCD}$.
We neglect lepton masses throughout this paper.

Besides the expression shown in (\ref{hcpc}), which is antisymmetric in
the neutrino momenta, the graph of Fig.~\ref{figwbox} also yields terms
symmetric in $q_1$ and $q_2$. These contributions cannot induce
$K_L\to\pi^0\nu\bar\nu$ in the limit of CP conservation. A similar 
comment holds for the $Z$-penguin diagram, which is entirely symmetric 
in $q_1$ and $q_2$. In principle also intermediate top quarks could
yield an interaction analogous to (\ref{hcpc}), but this contribution
is strongly suppressed by small CKM couplings.

Parametrically the amplitude arising from (\ref{hcpc}) is suppressed
with respect to the leading CP violating $K_L\to\pi^0\nu\bar\nu$
amplitude by a factor 
\begin{equation}\label{ome}
\delta^{SD}_{CPC} = {m^2_K \over M^2_W}
{\lambda_c\ln(m_c/\Lambda_{QCD}) \over {\rm Im}\lambda_t X_0(x_t)}
\sim {\cal O}(10\%)~. 
\end{equation}
Note that the very strong suppression due to the
smallness of $m^2_K/M^2_W$ is at least partially compensated by a large
enhancement factor $\lambda_c/{\rm Im}\lambda_t$, reflecting the CP
conserving nature of the mechanism under consideration.
Since the CP conserving amplitude is antisymmetric in the neutrino
momenta, there is no interference with the leading CP violating
contribution in the integrated rate. Therefore the CP conserving part
is simply to be added in rate (rather than amplitude) to the 
CP violating one. From the naive
order-of-magnitude estimate given above it would thus seem that an
effect in the percent range could still be possible. We will see in
the following that the actual size is in fact considerably smaller.

To evaluate the amplitude generated by the dimension-8 operator in
(\ref{hcpc}) one has to consider hadronic matrix elements of the
form
$\langle\pi^0|\bar s\Dleftad \gamma_\mu(1-\gamma_5)d|
K^0\rangle$,
involving quark currents with QCD covariant derivatives. We will next
determine these matrix elements to leading order in chiral
perturbation theory \cite{GL}. In this framework the lowest order 
realization of the QCD Lagrangian in the chiral limit, as a 
function of light pseudoscalar fields and external sources, 
is given by 
\begin{equation}\label{ls2}
{\cal L}^{(2)}_S=\frac{f^2}{8} {\rm tr}\left\{
\DUmu U \DDmu U^\dagger +\chi U^\dagger+ U\chi^\dagger
\right\}~,
\end{equation}
where $U$, transforming as $U\to g_R U g^\dagger_L$   
under $SU(3)_R\otimes SU(3)_L$ chiral rotations,
can be written as $U=\exp(2i\Phi/f)$ with
\begin{equation}\label{phi}
\Phi=\left[
\begin{array}{ccc}
\frac{\pi^0}{\sqrt{2}}+\frac{\eta}{\sqrt{6}} & \pi^+ & K^+ \\
\pi^- & -\frac{\pi^0}{\sqrt{2}}+\frac{\eta}{\sqrt{6}} & K^0 \\
K^- & \bar K^0 & -\frac{2\eta}{\sqrt{6}} \\
\end{array}\right]~.
\end{equation}
Here $f=132\, {\rm MeV}$ is the pion decay constant, 
$\chi$, transforming as $U$, is the external scalar source 
and $\DDmu = \partial_\mu U-i r_\mu U + i U l_\mu$
is the $SU(3)_R\otimes SU(3)_L$ covariant derivative 
which depends on the vector  sources $r_\mu$ and $l_\mu$
(see e.g. \cite{DAI} for conventions and a review on the subject). 
As usual, to incorporate mass terms we shall replace
\begin{equation}\label{chim}
\chi\to 2 B M~,\qquad\qquad  M={\rm diag}(m_u, m_d, m_s)~,
\end{equation}
where $B$ is a real constant that can be expressed in terms 
of quark and meson masses (e.g. to lowest order 
$m^2_{K^0}=B(m_s+m_d)$).
 
The derivatives of ${\cal L}^{(2)}_S$ with respect to the
external sources allow us to compute the 
lowest-order matrix elements of the corresponding quark currents.
For instance, this leads to
\begin{eqnarray}\label{vasp}
\bar q_i\gamma_\mu(1-\gamma_5)q_j 
&=& i \frac{f^2}{2} (\partial_\mu U^\dagger U)_{ji} 
+O(p^3) ~,\\ 
\bar q_i(1-\gamma_5)q_j 
&=&  -\frac{f^2}{2}B U^\dagger_{ji} 
+O(p^2)~,
\end{eqnarray}
where the above equalities have to be understood as 
identities of the corresponding matrix elements.
For the derivative operators we are interested in here, 
the lowest non-vanishing order is ${\cal O}(p^2)$ and
this statement alone would lead to the $m^2_K$ factor in the 
naive estimate (\ref{ome}). To be more specific,
chiral symmetry implies the following 
general representation at  ${\cal O}(p^2)$ 
\begin{eqnarray}\label{qdq1}
\bar q_i \Dleftad\gamma_\mu(1-\gamma_5)q_j &=&
i[g_{\alpha\mu}(c_0 \chi^\dagger U+\bar c_0 U^\dagger\chi) 
+c_1\partial_\alpha U^\dagger\partial_\mu U+
c_2\partial_\mu U^\dagger\partial_\alpha U \nonumber \\
&&+c_3\partial_\alpha\partial_\mu U^\dagger U-
i c_4\varepsilon_{\alpha\mu\beta\nu}\partial^\beta U^\dagger
\partial^\nu U]_{ji}~.
\end{eqnarray}
Note that the further possible term 
$U^\dagger\partial_\alpha\partial_\mu U$
satisfies the identity
\begin{equation}\label{uid}
U^\dagger\partial_\alpha\partial_\mu U\equiv
-\partial_\alpha\partial_\mu U^\dagger U-
\partial_\alpha U^\dagger\partial_\mu U-
\partial_\mu U^\dagger\partial_\alpha U 
\end{equation}
and is not independent of those already present in (\ref{qdq1}).

CP invariance of the strong interactions implies that all coefficients
$c_0$, $\bar c_0$, $c_1$, $\ldots$, $c_4$ are real (we employ the
CP convention $CP\, \Phi_{ij}=-\Phi_{ji}$). Adding to (\ref{qdq1}) its
hermitian conjugate yields an expression for
$\partial_\alpha(\bar q_i\gamma_\mu(1-\gamma_5)q_j)$
in terms of the $c_i$. This can be compared with the derivative of the
$V-A$ current from (\ref{vasp}). Taking into account (\ref{uid}), we
obtain
\begin{equation}\label{c0213}
c_0=\bar c_0~,\qquad c_2-c_1=\frac{f^2}{4}\qquad{\rm and}\qquad
c_3=\frac{f^2}{4}~.
\end{equation}
Additional constraints follow from the quark equations of motion
\begin{equation}\label{qeom}
\bar q_i\not\!\!{\overleftarrow D}(1-\gamma_5)q_j=
i\bar q_i m_i(1-\gamma_5)q_j=-i\frac{f^2}{4}(U^\dagger\chi)_{ji}~,
\end{equation}
or equivalently the equations of motion of the $U$ field
\begin{equation}\label{ueom}
(\partial^2 U^\dagger U-U^\dagger \partial^2 U)_{ji}=
(\chi^\dagger U-U^\dagger\chi)_{ji}~, \qquad\quad (i\not=j)~.
\end{equation}
Comparing (\ref{qeom}) with (\ref{qdq1}) for $\mu=\alpha$ and using
\begin{eqnarray}\label{duid}
\partial^2 U^\dagger U &\equiv&
\frac{1}{2}(\partial^2 U^\dagger U+U^\dagger\partial^2 U)+
\frac{1}{2}(\partial^2 U^\dagger U-U^\dagger\partial^2 U) \\ &=&
-\partial U^\dagger \partial U+
\frac{1}{2}(\chi^\dagger U-U^\dagger\chi)~, \qquad\quad
\end{eqnarray}
one finds
\begin{equation}\label{c0312}
c_0=-\frac{1}{8}c_3\qquad{\rm and}\qquad c_1+c_2=\frac{f^2}{4}~.
\end{equation}
Together with (\ref{c0213}) we then have
\begin{eqnarray}\label{qdq2}
\bar q_i \Dleftad \gamma_\mu(1-\gamma_5)q_j &=&
i\frac{f^2}{4}[\partial_\mu U^\dagger\partial_\alpha U+
\partial_\alpha\partial_\mu U^\dagger U-
\frac{1}{8}g_{\alpha\mu}(\chi^\dagger U+U^\dagger\chi)]_{ji}
\nonumber \\
&& +c_4\varepsilon_{\alpha\mu\beta\nu}(\partial^\beta U^\dagger
\partial^\nu U)_{ji}~.
\end{eqnarray}
Recalling that in our convention $CP |K\rangle =-|\bar K\rangle$ 
and $|K_L\rangle=(|K\rangle+|\bar K\rangle)/\sqrt{2}$, 
we finally get
\begin{equation}\label{mefin}
\langle\pi^0(p)|T_{\alpha\mu}|K_L(k)\rangle=
-\frac{i}{2}[(k-p)_\alpha (k+p)_\mu+\frac{1}{4}m^2_K g_{\alpha\mu}]~,
\end{equation}
where the undetermined coefficient $c_4$ has dropped out.

Interestingly enough, 
the matrix element (\ref{mefin}) gives zero when multiplied with the
leptonic current from (\ref{hcpc}). 
We then conclude that
$\langle\pi^0\nu\bar\nu|{\cal H}_{CPC}|K_L\rangle$
vanishes to leading order in chiral perturbation theory and the CP
conserving $K_L\to\pi^0\nu\bar\nu$ transition from ${\cal H}_{CPC}$
therefore receives an additional ${\cal O}(m_K^2/(8\pi^2 f^2))$
suppression.

To obtain a quantitative estimate we may write
\begin{equation}\label{meac}
\langle\pi^0(p)|T_{\alpha\mu}|K_L(k)\rangle=
i\frac{a_\chi}{2}(k+p)_\alpha (k+p)_\mu~,
\qquad a_\chi={\cal O}\left(\frac{m^2_K}{8\pi^2 f^2}\right)
\sim 20\%~.
\end{equation}
Introducing the kinematical variables
\begin{equation}\label{y12}
y_{1,2}=\frac{2 k\cdot q_{1,2}}{m^2_K}~,\qquad y=\frac{y_1-y_2}{2}~,
\end{equation}
\begin{equation}\label{uz}
u=\frac{(q_1+q_2)^2}{m_K^2}\quad{\rm and}\quad
z=\frac{m^2_\pi}{m^2_K}
\end{equation}
(with $q_1$ ($q_2$) the (anti)neutrino momentum), this implies
\begin{equation}\label{asdcpc}
\left| A(K_L\to\pi^0\nu\bar\nu)_{CPC}^{SD}\right| = \left|
y~ a_\chi~ \delta_{CPC}^{SD}~ A(K_L\to\pi^0\nu\bar\nu)_{CPV}\right|.
\end{equation}
Including the phase space integrations we obtain for the decay rates
\begin{equation}\label{gsdcpc}
\Gamma({K_L\to\pi^0\nu\bar\nu})_{CPC}^{SD} = 
\Gamma({K_L\to\pi^0\nu\bar\nu})_{CPV}\,
\left|a_\chi\right|^2\left|\delta_{CPC}^{SD}\right|^2 R_{kin}~,
\end{equation}
where ($\lambda(a,b,c)=a^2+b^2+c^2-2ab-2ac-2bc$)
\begin{equation}\label{rkin}
R_{kin} = \dis\frac{ \int{\rm d}\Gamma_{\pi^0\nu\bar\nu}~
[\lambda(1,u,z)-4y^2]y^2  }{ \int 
{\rm d}\Gamma_{\pi^0\nu\bar\nu}~[\lambda(1,u,z)-4y^2]} 
=\frac{1}{30}\frac{g(z)}{f(z)}~\simeq~0.03
\end{equation}
and 
\begin{eqnarray}
g(z) &=& 1-9z+45 z^2-45 z^4+9 z^5-z^6+60 z^3 \ln z~,\\
f(z) &=& 1-8z+8 z^3-z^4-12 z^2 \ln z~.
\end{eqnarray}
We thus find that the CP conserving rate
$\Gamma({K_L\to\pi^0\nu\bar\nu})_{CPC}^{SD}$ is further suppressed
by phase space, in addition to the parametric effect from
$\delta^{SD}_{CPC}$ and the chiral suppression described by $a_\chi$.
Numerically the three suppression factors on the rhs of (\ref{gsdcpc})
give 
$\sim 4\cdot 10^{-2}\times 10^{-2}\times 3\cdot 10^{-2}= 10^{-5}$.
We therefore conclude that $\Gamma({K_L\to\pi^0\nu\bar\nu})_{CPC}^{SD}$
is safely negligible by a comfortably large margin.

\section{The long-distance part of the CP conserving amplitude}

\begin{figure}[t]
    \begin{center}
       \setlength{\unitlength}{1truecm}
       \begin{picture}(7.0,3.5)
       \epsfxsize 7.0 true cm
       \epsfysize 3.5 true cm
       \epsffile{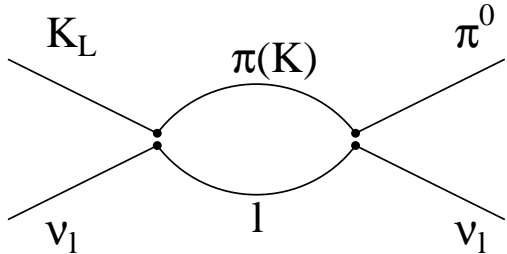}
       \end{picture} 
    \end{center}
    \caption{Long-distance diagram which contributes to the 
CP conserving $K_L\to\pi^0\nu\bar\nu$ amplitude.}
    \protect\label{figwbox2}
\end{figure}

Chiral perturbation theory provides a reliable framework
also to estimate the long-distance CP conserving contribution to 
$K_L\to\pi^0\nu\bar\nu$ generated by the up-quark exchange
in Fig.~\ref{figwbox}. The lowest-order contributions 
correspond to the charged $\pi$ and $K$ exchange in 
Fig.~\ref{figwbox2}, added to an appropriate local counterterm.
Since the lowest-order coupling of the pseudoscalar mesons 
with the leptonic currents can be derived from 
${\cal L}^{(2)}_S$, the loop amplitude is completely determined. 
The result is logarithmically divergent and is given by
\begin{equation}\label{ALcpc}
A(K_L\to\pi^0\nu\bar\nu)_{CPC}^{LD} = 
\frac{G_F}{\sqrt{2}}\frac{\alpha}{2\pi\sin^2\Theta_W}
\lambda_u~{1\over M_W^2}~H_\mu \bar\nu\gamma^\mu(1-\gamma_5)\nu~,
\end{equation}
where
\begin{equation}\label{hpik}
H_\mu=H^\pi_\mu+H^K_\mu~,
\end{equation}
\begin{eqnarray}
H^\pi_\mu &=& \frac{1}{2}k_\mu~ m^2_K y\left[\ln\frac{\mu^2}{m^2_K}+2+
\right. \nonumber \\
&&+\left.\frac{(1-y_1)\ln(1-y_1)-(1-y_2)\ln(1-y_2)}{y_1-y_2}+i\pi\right]~,
\label{hpi}\\
H^K_\mu &=& \frac{1}{4}k_\mu~ m^2_K y\left[\ln\frac{\mu^2}{m^2_K}+2+
\frac{1}{y_1-y_2}\left(\frac{y^2_1\ln y_1}{1-y_1}-
\frac{y^2_2\ln y_2}{1-y_2}\right)\right]~.\label{hka}
\end{eqnarray}
Here we have used dimensional regularization and subtracted the
divergence according to the $\overline{MS}$ prescription. 
The imaginary part is due to the absorptive contribution of the pion.
We are neglecting the pion mass in the loop integration.
The $\mu$ dependence in (\ref{hpi}-\ref{hka}) has to be canceled 
by the coun\-ter\-term, whose finite contribution is however not know.
As a matter of fact, adding the counterterm to the loop result
would just produce the effect of fixing $\mu$ in 
(\ref{hpi}-\ref{hka}) to some typical hadronic scale.

In principle, looking at the quark level result, 
one could think that the $\mu$ dependence of 
(\ref{hpi}-\ref{hka}) should be partially related to
the scale dependence of the charm 
contribution in (\ref{hcpc}). However, the vanishing 
of $\langle\pi^0\nu\bar\nu|{\cal H}_{CPC}|K_L\rangle$ 
to the lowest order implies that there is no direct matching between
the two contributions at this level: 
the ultraviolet cut-off of the leading 
long-distance contribution is more likely 
to be a hadronic scale below the charm mass.

An indirect confirmation of the above statement is obtained
estimating higher order contributions in the framework
of vector meson dominance. In this limit the point-like
form factors of the  vector currents
in Fig.~\ref{figwbox2} are replaced by vector meson 
propagators ($1\to M_V^2/(M_V^2 - q^2)$) and
as a result the loop amplitude becomes finite. Note that
other potentially divergent contributions
generated by vector or axial-vector exchange
vanish to leading order, as can be checked explicitly 
using the lowest-order chiral Lagrangian of \cite{EGPR}.
Thus the main effect of vector resonances is just to 
provide a natural cut-off for the loop amplitude of 
Fig.~\ref{figwbox2}. Expanding the full result thus 
obtained in $1/M_V$ and neglecting terms suppressed by powers of
$m^2_K/M^2_V$, one recovers an amplitude of the form 
(\ref{hpi}-\ref{hka}) with
\begin{equation}\label{mumv}
\ln\frac{\mu^2}{m^2_K}\to \ln\frac{M^2_V}{m^2_K}-\frac{3}{2}~.
\end{equation} 
It is convenient to further expand the result in the kinematical
variables $y$ and $u$, which yields
\begin{equation}\label{Hmu2}
H_\mu = \frac{1}{2}k_\mu~ m_K^2 y \left[\frac{3}{2} 
\ln\frac{M^2_V}{m^2_K}+\frac{1}{4}-\frac{1}{2}\ln 2+i\pi
+{\cal O}\left(u,y^2\right) \right]~.
\end{equation}
Comparing (\ref{Hmu2}) with (\ref{hpi}-\ref{hka}) it is seen that the
common vector meson mass $M_V\simeq 800$~MeV (we can safely 
neglect the small $SU(3)$ breaking effects in the vector meson sector)
provides the ultraviolet cut-off for the lowest-order calculation. 
The expansion in powers of the 
kinematical variables is well justified not only by the 
size of  the higher order terms but, especially,
because the phase-space integration strongly suppresses 
their contribution to the total rate. 

Using the approximate expression (\ref{Hmu2})
the integration over the $\pi^0\nu\bar\nu$ Dalitz plot
can be done analytically. Normalizing the CP conserving 
amplitude to the leading CP violating term, generated by 
(\ref{heff}), we can write
\begin{equation}
\left| A(K_L\to\pi^0\nu\bar\nu)_{CPC}^{LD}\right| = 
\left|y~\delta_{CPC}^{LD}~ A(K_L\to\pi^0\nu\bar\nu)_{CPV}\right|,
\end{equation}
where, analogously to (\ref{ome}), we have defined 
\begin{equation}\label{omeld}
\delta^{LD}_{CPC} =\frac{m^2_K}{4 M^2_W}
{\lambda_u \over {\rm Im}\lambda_t X_0(x_t)} 
\left| \frac{3}{2}\ln\frac{M^2_V}{m^2_K}+\frac{1}{4}-\frac{1}{2}\ln 2
+i\pi \right|~\approx~0.04~.
\end{equation}
Taking into account the phase space integration,
the CP conserving rate can be written as
\begin{equation}\label{ldcpc}
\Gamma({K_L\to\pi^0\nu\bar\nu})_{CPC}^{LD} = 
\Gamma({K_L\to\pi^0\nu\bar\nu})_{CPV}\,
\left|\delta_{CPC}^{LD}\right|^2 R_{kin}~.
\end{equation}
The total suppression factor on the rhs of (\ref{ldcpc}) is thus
estimated to be 
$\sim 2\cdot 10^{-3}\times 3\cdot 10^{-2}=6\cdot 10^{-5}$.

\section{Conclusions}

Even if CP would be exactly conserved, the decay $K_L\to\pi^0\nu\bar\nu$
could in principle proceed within the Standard Model. We have 
presented a detailed analysis of both the short-distance and the
long-distance contribution to the CP conserving rate. We have shown
that these contributions are suppressed by more than four orders of
magnitude in comparison to the leading, direct CP violating
branching ratio of about $3\cdot 10^{-11}$.
Several reasons are responsible for this very strong suppression:
first, the CP conserving amplitude, which is parametrically 
${\cal O}(10\%)$ of the direct CP violating one, does not interfere
with the latter, but is added simply in rate. In addition, there is a
substantial suppression factor ($\sim 0.03$) from phase space. Finally,
the short-distance part of the CP conserving amplitude
is also chirally suppressed.

The hierarchy of the direct CP violating, indirect CP violating
\cite{BB6} and CP conserving contributions to $B(K_L\to\pi^0\nu\bar\nu)$
in the Standard Model is thus 
$1~:~10^{-2}~:~
\;\raisebox{-.4ex}{\rlap{$\sim$}} \raisebox{.4ex}{$<$}\; 10^{-4}$.
Note however that in the absence of direct CP violation, as in a 
superweak model \cite{WW}, the $K_L\to\pi^0\nu\bar\nu$ branching
fraction due to indirect CP violation alone would only be
$\sim 2\cdot 10^{-15}$, comparable to $B(K_L\to\pi^0\nu\bar\nu)_{CPC}$.
The above hierarchy of contributions can be contrasted with the case of
$K_L\to\pi^0e^+e^-$, where all three mechanisms are of roughly
comparable magnitude \cite{DG}.

We have also found that the calculation of the CP conserving 
$K_L\to\pi^0\nu\bar\nu$ rate is quite well under control, although
the precise value depends on long-distance dynamics that is hard to
quantify in detail. The extremely strong suppression of such
contributions highlights the theoretically clean nature of
$K_L\to\pi^0\nu\bar\nu$ in a striking manner.

\section*{Acknowledgments}

We thank Martin Beneke and Gilberto Colangelo for discussions.
This work has been partially supported by
the EEC-TMR Program, Contract N. CT98-0169.

\vfill\eject

\end{document}